\documentclass[aps,prl,preprint,superscriptaddress,showpacs,preprintnumbers,amsmath,amssymb]{revtex4}
\usepackage{graphicx}

\usepackage[usenames]{color}
\graphicspath{{ps/}}

\begin{document}

\title{\boldmath $CP$ violation in {$b\to c\bar cd$} decays at Belle}

\author{T.~Aushev}

\address{for the Belle Collaboration\\
  ITEP, B.~Cheremushkinskaya, 25, Moscow, Russia\\
  E-mail: aushev@itep.ru}

\begin{abstract}
  We report a measurement of the $CP$ violation parameters in $b\to
  c\bar cd$ decays.  Mixing-induced $CP$ violation involving this
  transition is studied using a sample of $B^0\to J/\psi\pi^0,
  D^{*+}D^{*-}$ and $D^{*\pm}D^\mp$ decays with a data set accumulated
  at the $\Upsilon(4S)$ resonance with the Belle detector at the KEKB
  energy-asymmetric $e^+e^-$ collider.  $CP$ violation parameters are
  extracted from a fit to the distributions of time intervals between
  two $B^0$ meson decay vertexes.
\end{abstract}

\pacs{11.30.Er, 12.15.Hh, 13.25.Hw}
\maketitle

\section{Introduction}

In the Standard Model (SM), $CP$ violation arises from the
Kobayashi-Maskawa (KM) phase in the weak interaction quark-mixing
matrix.  The SM predicts $CP$ asymmetries in the time-dependent rates
for $B^0$ and $\bar B^0$ decays to a common $CP$ eigenstate $f_{CP}$.
Recent measurements of the $CP$ violation parameter $\sin2\phi_1$ by
Belle\cite{belle} and BaBar\cite{babar} collaborations established
$CP$ violation in $B^0\to J/\psi K_S$ and related decay modes at a
level consistent with KM expectations.  Comparisons between SM
expectations and measurements in various modes are important to test
the KM model.  In particular, the $CP$ violation in the decays,
dominated by the tree $b\to c\bar cd$ transition, is sensitive to the
same angle $\phi_1$ as in the $b\to c\bar cs$.  If other contributions
to the same final state are substantial, a precision measurement of
the time-dependent $CP$ asymmetry in $b\to c\bar cd$ may reveal a
phase that differs from $\phi_1$.  One of the contribution leading to a
different weak phase, is a penguin diagram, which is expected to be
small in the SM.  Thus, the measurements of $CP$ asymmetries in $b\to
c\bar cd$ transitions play an important role to probe the SM.

The analyses described here are based on $140\,$fb$^{-1}$ of data,
corresponding to $152$ million $B\bar B$ pairs, collected with the
Belle detector at the KEKB asymmetric energy storage rings.

\section{\boldmath Study of the $B^0\to J/\psi\pi^0$}

\begin{figure}[htb]
  \includegraphics[width=.7\textwidth]{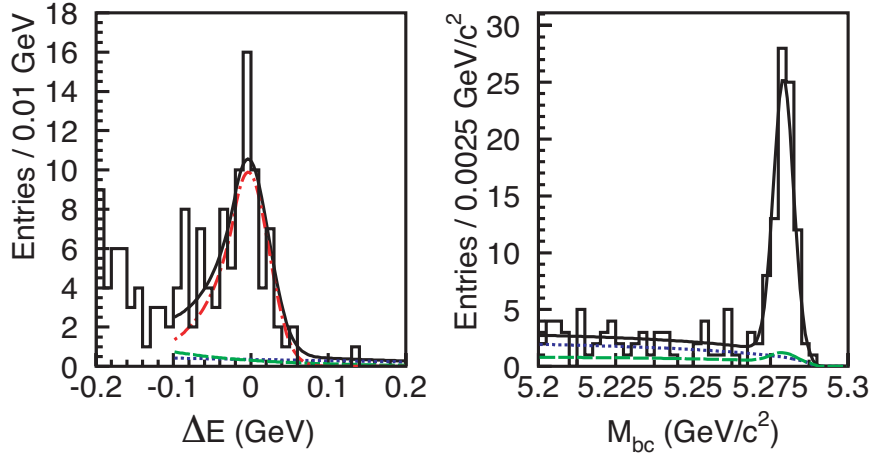}
  \caption{$\Delta E$ and $M_{bc}$ distributions for $B^0\to
    J/\psi\pi^0$ candidates.  The superimposed curves show fitted
    contribution from signal (dot-dashed line), $B\to J/\psi X$
    background (dotted line), combinatorial background (dashed line)
    and the sum of all the contributions (solid line).}
  \label{jpsipi0sig}
\end{figure}

The details of the analysis are described in Ref.\cite{jpsipi0}.  $B$
meson candidates are identified using the beam constrained mass
$M_{bc}\equiv\sqrt{E_{beam}^2-P_B^2}$ and the energy difference
$\Delta E\equiv E_B-E_{beam}$, where $E_B$($P_B$) is the
energy(momentum) of the $B$ candidate and $E_{beam}$ is the
center-of-mass (CM) beam energy.  For extracting the $CP$ asymmetry,
the standard Belle tagging and vertexing procedures\cite{belle} are
used.  After flavor tagging and vertex reconstruction, 91 $B^0\to
J/\psi\pi^0$ candidates with $(84\pm11)\%$ purity are selected.
$\Delta E$ and $M_{bc}$ distributions for the candidate events are
shown in Fig.~\ref{jpsipi0sig}.

In the decay chain $\Upsilon(4S)\to B^0\bar B^0\to f_{CP}f_{tag}$,
where one of the $B$ meson decays at time $t_{CP}$ to a final state
$f_{CP}$ and the other decays at time $t_{tag}$ to a final state
$f_{tag}$ that distinguishes between $B^0$ and $\bar B^0$, the decay
rate has a time dependence given by\cite{cprate}
\begin{eqnarray} 
  {\cal P}(\Delta t)= 
  \frac{e^{-|\Delta t|/\tau_{B^0}}}{4\tau_{B^0}}
  \{1+q[S_{f_{CP}}\sin(\Delta m_d\Delta t)
    +C_{f_{CP}}\cos(\Delta m_d\Delta t)]\},
  \label{eq:cprate}
\end{eqnarray}
where $\tau_{B^0}$ is the $B^0$ lifetime, $\Delta m_d$ is the mass
difference between the two $B^0$ mass eigenstates, $\Delta t\equiv
t_{CP}-t_{tag}$, and the $b$-flavor $q=+1(-1)$ when the tagging $B$
meson is a $B^0$($\bar B^0$).  In $b\to c\bar cd$ transition the SM
predicts $S_{f_{CP}}=-\xi_f\sin2\phi_1$ and $C_{f_{CP}}=0$, where
$\xi_f=+1(-1)$ corresponds to $CP$-even (-odd) final state ($\xi_f=+1$
in case of $J/\psi\pi^0$).

\begin{figure}[htb]
  \includegraphics[width=.5\textwidth]{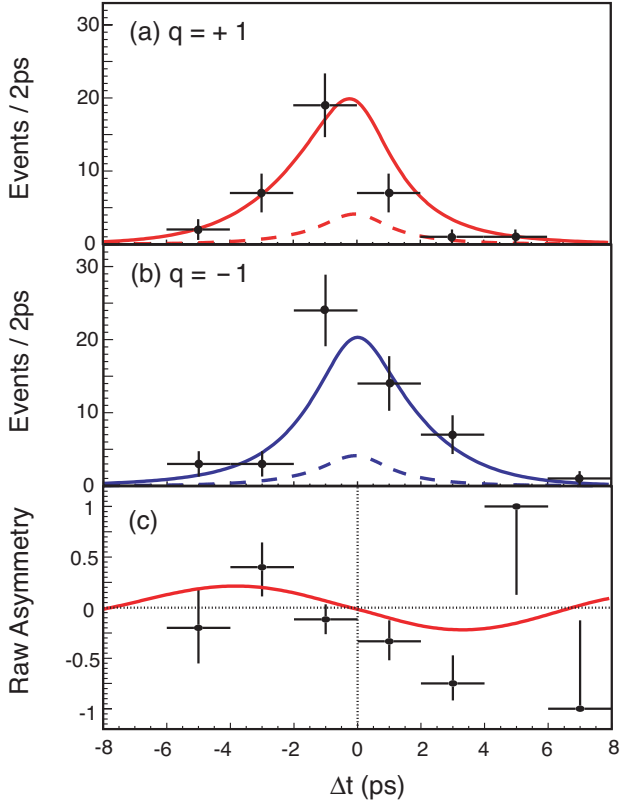}
  \caption{The $\Delta t$ distributions for a) $B^0\to J/\psi\pi^0$
    ($q=+1$) and b) $B^0\to J/\psi\pi^0$ ($q=-1$) and the raw
    asymmetry c) for the good tag region.  The curves show the result
    of the fit, and dashed curves show the background distributions.}
  \label{jpsipi0dt}
\end{figure}

We determine $S_{f_{CP}}$ and $C_{f_{CP}}$ by performing an unbinned
maximum-likelihood fit to the observed $\Delta t$ distribution.  The
probability density function expected for the signal distribution is
given by Eq.~\ref{eq:cprate} taking into account the effect of
incorrect flavor assignment and the finite vertex resolution.  The fit
yields $CP$ violating parameters presented in Table~\ref{cpresult}.
Fig.~\ref{jpsipi0dt} shows $\Delta t$ distributions for $B\to
J/\psi\pi^0$ decays tagged as $B^0$ and $\bar B^0$ and $CP$ violation
asymmetry.

\section{\boldmath Study of the $B^0\to D^{*+}D^{*-}$}

\begin{figure}[htb]
  \begin{tabular}{@{\hspace{-.1cm}}l@{\hspace{-.5cm}}r}
    \includegraphics[width=.4\textwidth]{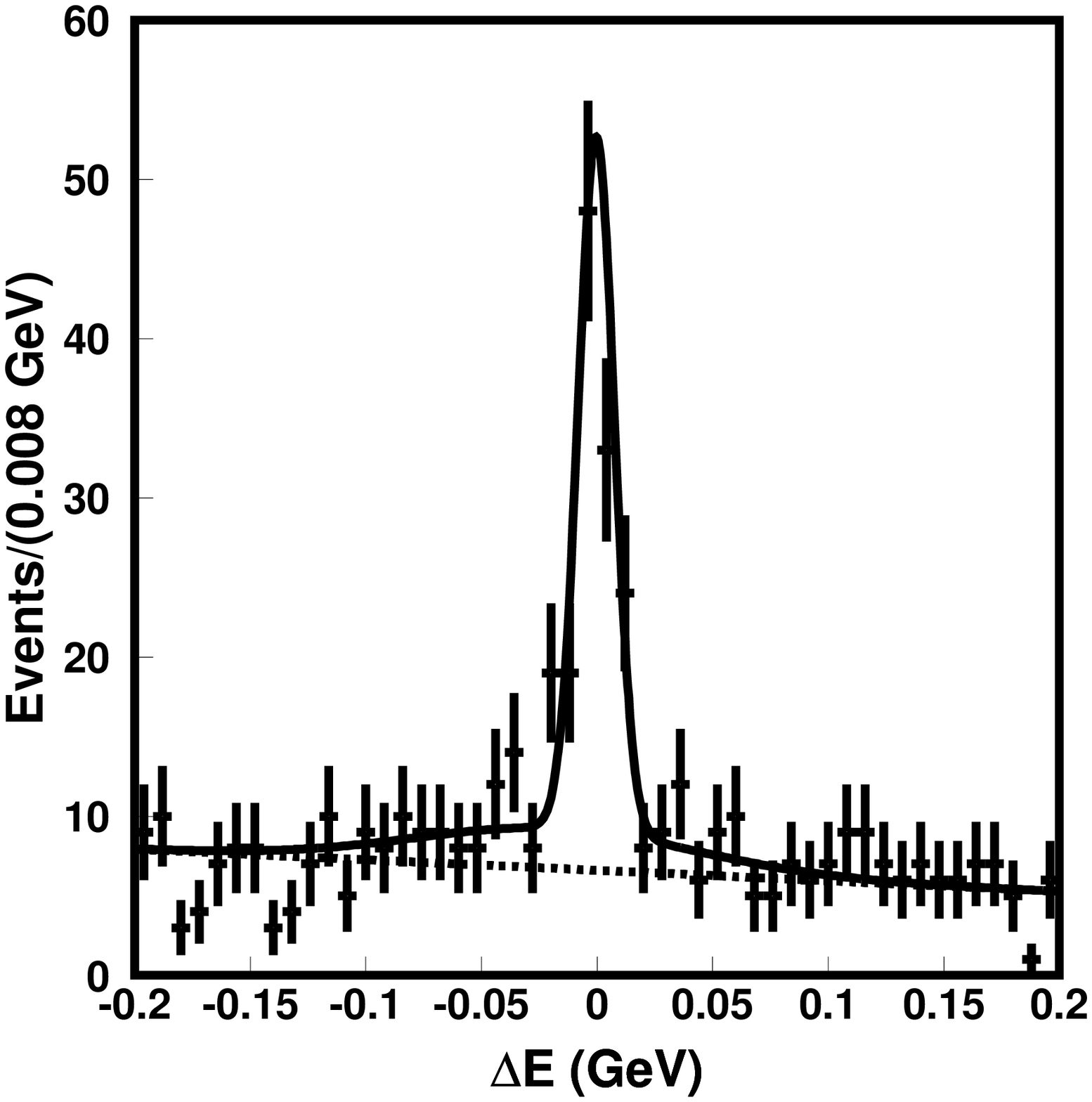} &
    \put(-130,145){a)}
    \includegraphics[width=.4\textwidth]{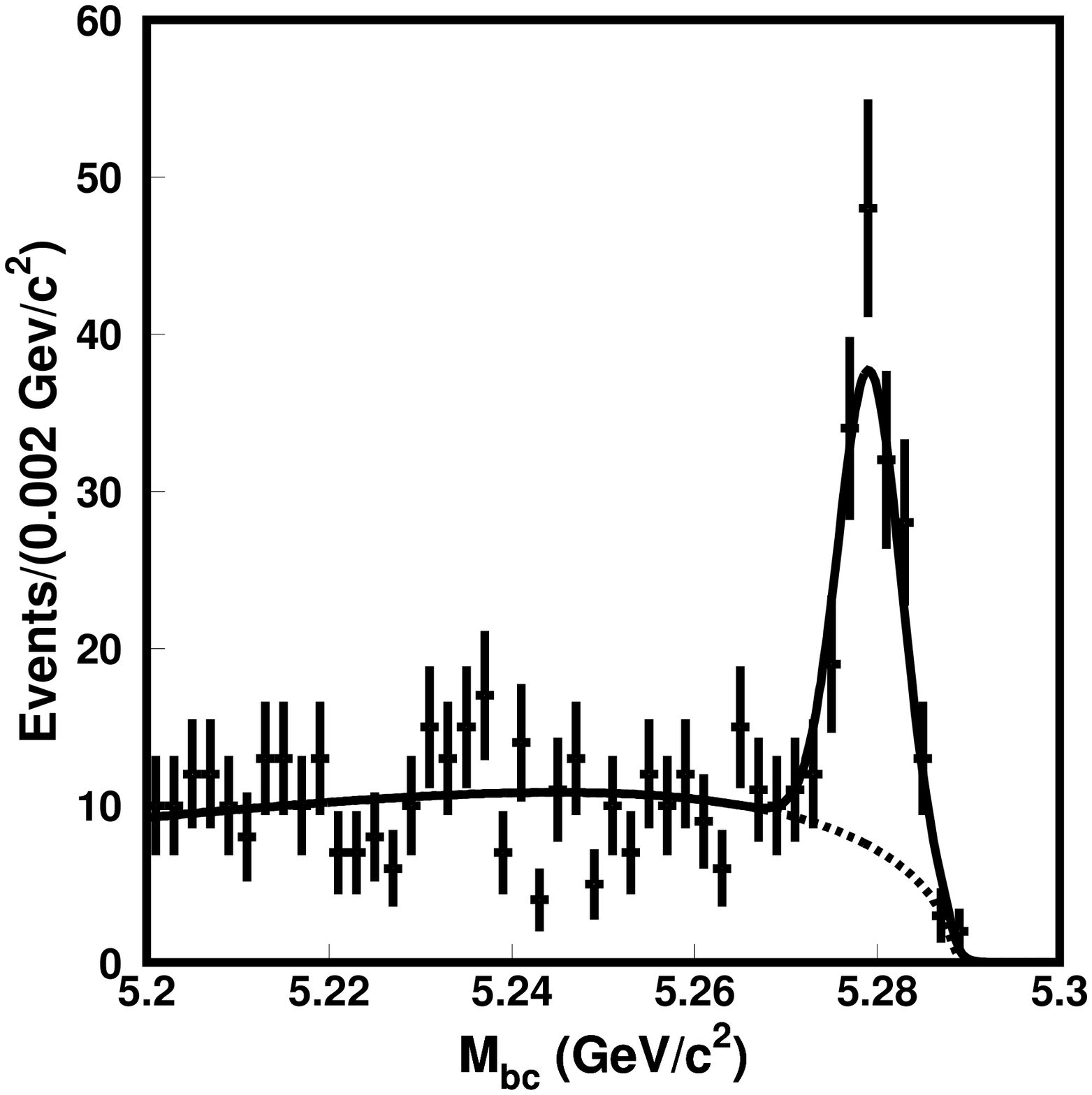}
    \put(-145,145){b)}
  \end{tabular}
  \caption{a) $\Delta E$ and b) $M_{bc}$ distributions for $B^0\to
    J/\psi\pi^0$ candidates.  Solid curves show the fit to signal plus
    background distributions, and dashed curves show the background
    contributions.}
  \label{dstdst_sig}
\end{figure}

$B^0\to D^{*+}D^{*-}$ meson candidates are identified using $\Delta E$
and $M_{bc}$ variables.  These distributions for the selected events
are shown in Fig.~\ref{dstdst_sig}.  The fit yields $130\pm13$ signal
events and the branching fraction is calculated to be
$[0.81\pm0.08({\rm stat})\pm0.11({\rm syst})]\times10^{-3}$.

\begin{figure}[htb]
  \begin{tabular}{@{\hspace{-.4cm}}l@{\hspace{-.7cm}}r}
    \includegraphics[width=.4\textwidth]{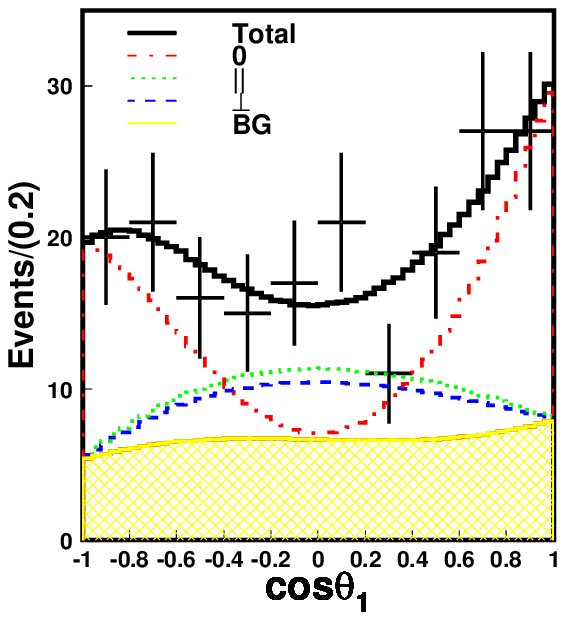} &
    \includegraphics[width=.4\textwidth]{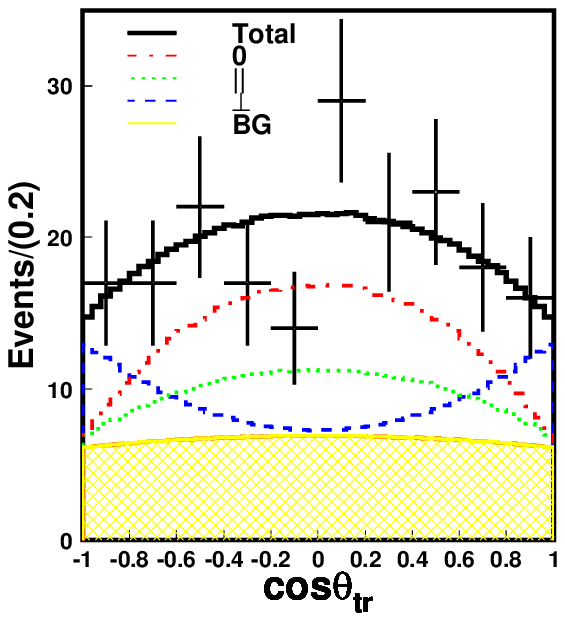}
  \end{tabular}
  \caption{The angular distributions of $B^0\to D^{*+}D^{*-}$
    candidates projected onto (left) $\cos\theta_{tr}$ and (right)
    $\cos\theta_1$.  The solid curves show the background, the
    dashed-dotted lines show $R_0$ polarization, the dotted lines show
    $R_\parallel$, the dashed lines show $R_\perp$ polarization.}
  \label{dstdst_pol}
\end{figure}

In the $B^0\to D^{*+}D^{*-}$ decay, the final state consists of S, P
and D waves, corresponding to different $CP$ eigenvalues, thus the
$CP$ violation parameters are diluted.  Corresponding fractions of
these final states $R_0$, $R_\parallel$ ($CP$-even) and $R_\perp$
($CP$-odd) are extracted from the transversity basis\cite{trans}
angular analysis using $\cos\theta_1$ and $\cos\theta_{tr}$
distributions (Fig.~\ref{dstdst_pol}).  The fit yields
$R_\perp=0.19\pm0.08({\rm stat})\pm0.01({\rm syst})$ and
$R_\parallel=0.57\pm0.08({\rm stat})\pm0.02({\rm syst})$.

\begin{figure}[htb]
  \includegraphics[width=.5\textwidth]{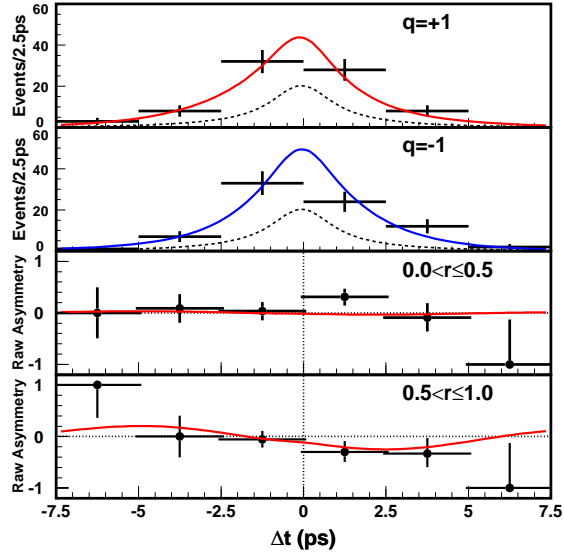}
  \caption{The $\Delta t$ distributions and the raw asymmetries for
    $B^0\to D^{*+}D^{*-}$ candidates.  The curves show the result of
    the fit, and dashed curves show the background distributions.}
  \label{dstdst_dt}
\end{figure}

We perform a simultaneous unbinned maximum likelihood fit to the
$\Delta t$, $\cos\theta_{tr}$ and $\cos\theta_1$ distributions for
$B^0\to D^{*+}D^{*-}$ candidates to measure $CP$ violation parameters.
The signal $B^0$ decay vertex is reconstructed by fitting $D$ mesons
momentum vectors with the constraint of the interaction region
profile.  The tagging probability and the $B$ vertices determination
are identical to our others $CP$ violation analysis\cite{belle}.  The
fit results are presented in Table~\ref{cpresult}.
Fig.~\ref{dstdst_dt} shows the $\Delta t$ distributions for $B^0\to
D^{*+}D^{*-}$ candidates and $CP$ violation asymmetry.

\section{\boldmath Study of the $B^0\to D^{*\pm}D^\mp$}

Although the $D^{*\pm}D^{\mp}$ final states are not $CP$ eigenstates,
they can be produced in the decays of both $B^0$ and $\bar B^0$ with
comparable amplitudes; the interference between amplitudes of the
direct transition and those via $B\bar B$ mixing results in $CP$
violation\cite{aleksan}.  The probability for a $B$ meson to decay to
$D^{*\pm}D^{\mp}$ at time $\Delta t$ can be expressed in terms of five
parameters, ${\cal A}$, $S_\pm$ and $C_\pm$:
\begin{eqnarray} 
  {\cal P}_{D^*D}^\pm(\Delta t)=(1\pm{\cal A})
  \frac{e^{-\left|\Delta t\right|/\tau_{B^0}}}{8\tau_{B^0}}
  \{1+q[S_\pm\sin(\Delta m_d\Delta t)-
    C_\pm\cos(\Delta m_d\Delta t)]\}.
  \label{babarparam}
\end{eqnarray}
Here the $+(-)$ sign represents the $D^{*+}D^-$ ($D^{*-}D^+$) final
state.

\begin{figure}[htb]
  \includegraphics[width=.7\textwidth]{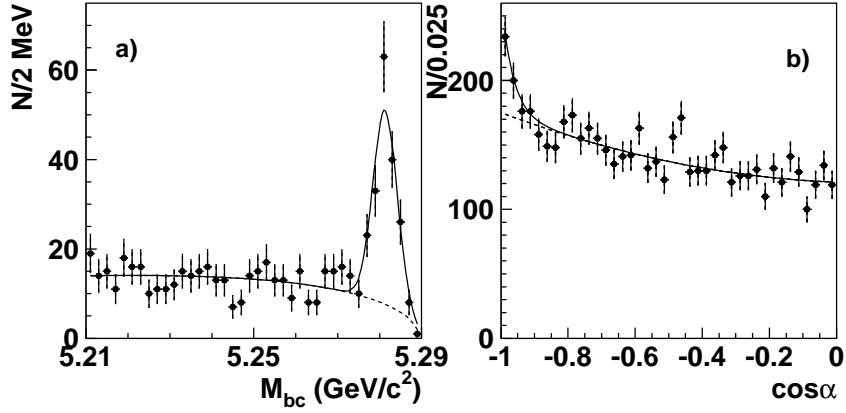}
  \caption{a) $M_{bc}$ and b) $\cos\alpha$ distributions of $B^0\to
    D^{*\pm}D^\mp$ candidates for the full and partial reconstruction
    methods, respectively.}
  \label{ddst_sig}
\end{figure}

Two reconstruction techniques, full and partial, are used to increase
the reconstruction efficiency\cite{ddstcp}.  In case of the full
reconstruction, a fit to the $M_{bc}$ distribution finds a signal
yield to be $161\pm16$ events.  In the partial reconstruction method
the angle $\alpha$ between the $D^-$ and $\pi_{slow}^+$ CMS momenta is
used to identify the studied decay.  In this case, we require the high
momentum lepton in the event to suppress the background and provide
the vertex and flavor of tagging $B$, while in the full reconstruction
method we use the standard tagging procedures\cite{belle}.  In both
cases, the signal $B^0$ vertex is reconstructed as in $B^0\to
D^{*+}D^{*-}$ analysis.  The fit to $\cos\alpha$ distribution yields
$137\pm39$ signal events, in good agreement with those expected from
the full reconstruction analysis.  Fig.~\ref{ddst_sig} shows the
signal distributions for selected candidates.

\begin{figure}[htb]
  \includegraphics[width=.5\textwidth]{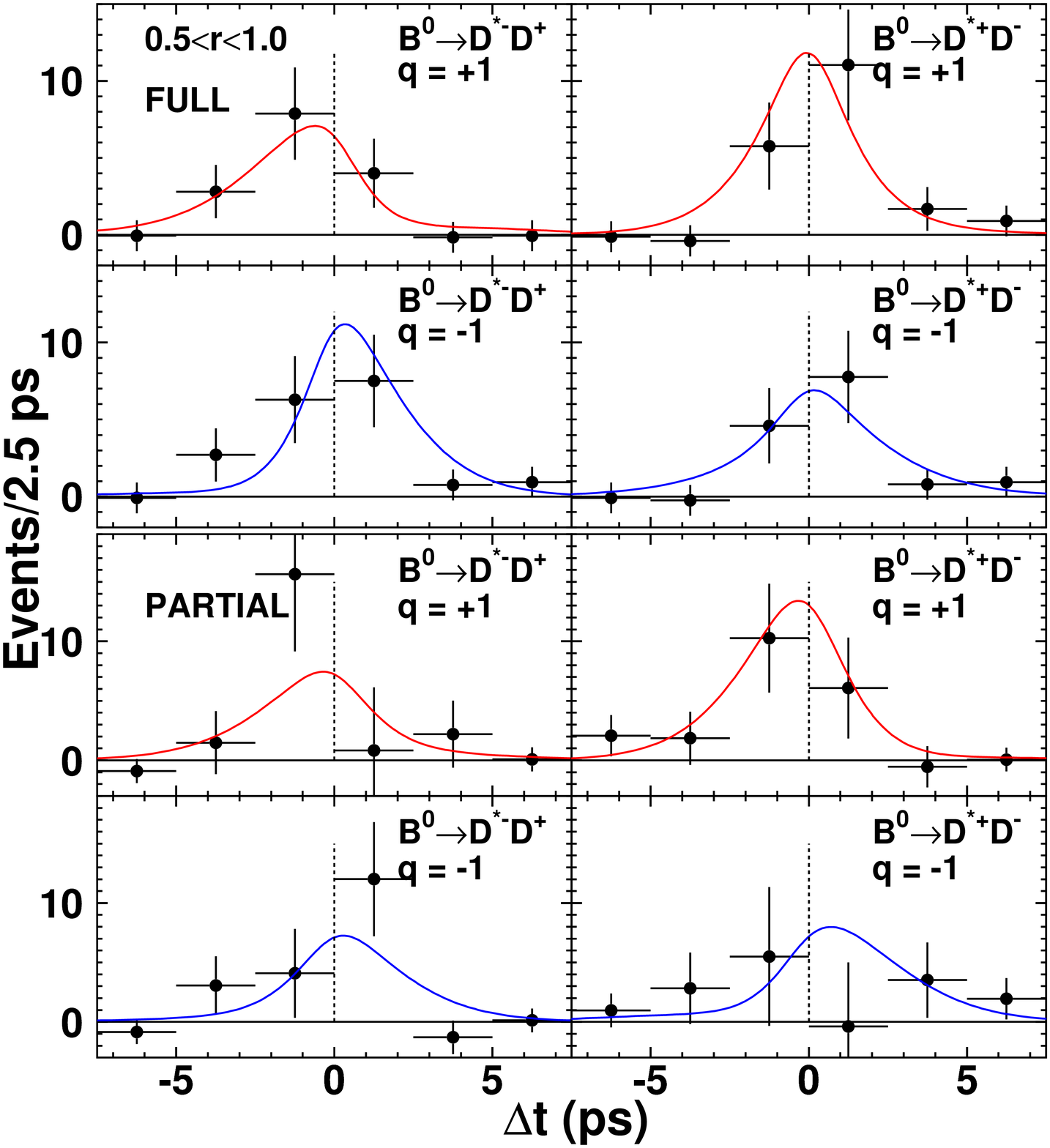}
  \caption{Background subtracted $\Delta t$ distributions in the full
    and partial reconstruction methods. The curves show the result of
    the fits.}
  \label{ddst_dt}
\end{figure}

The $\Delta t$ distributions after background subtraction are shown in
Fig.~\ref{ddst_dt} a) and b) for the full and partial reconstruction
methods, respectively.  The fit results are summarized in the
Table~\ref{cpresult}.

\section{Summary}

We measured $CP$ violation parameters in the $B^0\to J/\psi\pi^0,
D^{*+}D^{*-}$ and $D^{*\pm}D^\mp$ decays.  The results of these
measurements are presented in the Table~\ref{cpresult}, which are
consistent with the expectations from the SM.  We also report the
measurement of the branching fraction and the polarization parameters
for the $B^0\to D^{*+}D^{*-}$ decay mode.
\begin{table*}[htb]
  \begin{center}
    \caption{The summary for the $CP$ violating parameters
      measurement.}
    \label{cpresult}
    \begin{tabular}{|l|c|c|l|c|}
      \hline
      \raisebox{0pt}[16pt][6pt] 
      & $B^0\to J/\psi\pi^0$ & $B^0\to D^{*+}D^{*-}$
      & \multicolumn{2}{c|}{$B^0\to D^{*\pm}D^\mp$} \\
      \hline
      \raisebox{0pt}[16pt][6pt] 
      & & & ${\cal A}$ & $+0.07\pm0.08\pm0.04$ \\
      \hline
      \raisebox{0pt}[16pt][6pt]{$S_{f_{CP}}$}
      & $-0.72\pm0.42\pm0.09$ & $-0.75\pm0.56\pm0.12$ 
      & $S_-$ & $-0.96\pm0.43\pm0.12$ \\
      \cline{4-5}
      \raisebox{0pt}[16pt][6pt] & & & $S_+$ & $-0.55\pm0.39\pm0.12$ \\
      \hline
      \raisebox{0pt}[16pt][6pt]{$C_{f_{CP}}$}
      & $-0.01\pm0.29\pm0.03$ & $-0.26\pm0.26\pm0.04$ 
      & $C_-$ & $+0.23\pm0.25\pm0.07$ \\
      \cline{4-5}
      \raisebox{0pt}[16pt][6pt] & & & $C_+$ & $-0.37\pm0.22\pm0.07$ \\
      \hline
    \end{tabular}
  \end{center}
\end{table*}

This work is partially supported by Russian grant SS551722.2003.

\end{document}